\documentstyle[aps,preprint,epsf]{revtex}

\begin{document}
\draft
\preprint{US-FT/3-96}

\title{Self-organized criticality and the lattice topology}
\author {Alberto Saa}

\address{Departamento de F\'\i sica de Part\'\i culas\\
       Universidade de Santiago de Compostela\\
       15706 Santiago de Compostela, Spain}
\maketitle

\begin{abstract}
We examine exhaustively
the behavior of avalanches in 
critical height sandpile
models based in 
two- and three-dimensional
lattices of various topologies. 
We get that for two-dimensional lattices the spatial and temporal 
distributions characterizing
bulk avalanches do not depend on the lattice topology.
For the three-dimensional case, we detect a small dependence of
the topology for the temporal distribution, while the spatial ones
are independent.
The two-dimensional lattices studied are: the plane ($R^2$),
the cylinder ($S^1\times R$), and the M\"obius-strip ($M$); and the
three-dimensional are: $R^3$, $S^1\times R^2$, $S^1\times S^1\times R$, 
$M\times R$, $S^2\times R$,
$K\times R$, and $RP\times R$, where $K$ and $RP$ are respectively the
Klein bottle and the real projective plane.
\end{abstract}

\vspace{1cm}
\begin{flushleft}
\end{flushleft}

\newpage


The notion of self-organized criticality introduce by Bak, Tang, and
Wiesenfeld\cite{BTW}
 has been intensively used in the study of a huge class of phenomena
in recent years.
They showed that certain 
extended dissipative dynamical systems naturally evolve into a 
critical and stationary 
state, with no intrinsic length or time scale
and independent of initial conditions, through a
self-organization process. 
Their models are very simple simulations of real
avalanches in a sandpile\cite{N,P};
a cellular automaton in which the basic variable $h({\bf r})$ is
an integer describing the height or the slope
of a ``sandpile'' at the point $\bf r$
that belongs to an $n$-dimensional lattice. When $h({\bf r})$ exceeds
a critical value $k$ there is an avalanche and $h({\bf r})$ is
updated according to some automaton rules. 
This process is repeated until $h({\bf r})$ is less than $k$
for all points $\bf r$  in the lattice. 
Many numerical and analytical 
studies were based in these models and much
information is now available\cite{BT,an}.
The scale-invariant behavior automatically generated by the very
simple dynamics of these models is  the main motivation to 
the intensive activity of last years in the subject.
It is widely believed that the understanding of such
toy models can shed some light in the dynamics of real systems. For
instance, the original models of 
Bak, Tang, and Wiesenfeld were introduced  as an
intent to explain the ubiquitous in nature ``1/f'' noise\cite{DH}.

In \cite{BTW} is argued that near the self-organized critical
state the system has an universal behavior characterized by some
critical exponents. 
Evidences for universality in
the self-organized critical behavior of avalanches appeared first in
the work of Kadanoff, Nagel, Wu, and Zhou\cite{KNWZ}. They studied
one- and two-dimensional lattices of various sizes and their result
pointed toward the existence of certain universality classes.
For the two-dimensional
case they also considered models with various internal symmetries, and 
showed that finite-size effects in these models are rather subtle. 
Scalling properties and finite-size effects 
for various types of boundary conditions were also considered\cite{O,STC}.
Robustness is another crucial requirement necessary to the success of
these concepts in explaining real systems. 
The existence of self-organized critical states can not depend tightly
of especial features of the model.
In \cite{BTW} it is shown
that self-organized critical states in sandpile models 
are stable under some sort of quenched randomness.
The behavior of avalanches in a random lattice was considered in
\cite{P1}, and again one gets evidences for the existence of
a self-organized critical state, although with other critical exponents. 
Self-organized criticality  was also verified under lattice
cyclicities\cite{D}.

The dynamical process
of these models is {\em local} in essence; when a toppling occurs in a
given point only its nearest neighbor are affected. In spite of this, global
properties of the lattice can alter qualitatively the time evolution of
the system, as one can see, for instance, by comparing the number of
``sand grains'' that drop off the edges 
for lattices of different topologies. Indeed, one does not have {\em a priori}
motivations to expect that sandpile models based on lattices of distinct 
topologies evolve in the same way for large times. In
spite of the intense activity in this subject, the behavior of
avalanches in lattices with non-trivial topology has not been
considered yet. 

In this work we discuss the influence of global properties of the
lattice on the critical exponents of bulk avalanches. 
We consider two- and three-dimensional
lattices of distinct topologies and our result is rather
intriguing: for all cases we verify the existence of a
self-organized critical state, and we demonstrate, within our accuracy, 
that the critical exponents
for the spatial distributions characterizing avalanches
depend only on the dimension of the lattice and do not on its
particular topology. 
As to the temporal distributions, we detect a small dependence of the
critical exponents on the topology for three-dimensional lattices,
while for two-dimensional ones they are independent. 
Our result strongly 
corroborates the robustness conjectures
since we are dealing with lattices of very different global
structure; for instance, some of them are orientables and others are
not.  In our models, avalanches are characterized by cluster sizes ($s$), 
number of topplings ($m$), and by  fluctuation lifetimes ($t$). 
We have examined the following two-dimensional lattices:
the plane $R^2$,
the cylinder $S^1\times R$, and the M\"obius-strip $M$; and the
following 
three-dimensional ones: $R^3$, $S^1\times R^2$, $S^1\times S^1\times R$, 
$M\times R$, $S^2\times R$,
$K\times R$, and $RP\times R$, where $K$ and $RP$ are respectively the
Klein bottle and the real projective plane\cite{naka}. We simulate the
distinct topologies by imposing certain boundary conditions on the
standard lattice. Our result can be presented also as the fact that
such boundary conditions do not alter the behavior of avalanches. 
Incidentally, a recent work\cite{STC} also appreciates  the possibility
that the presence of closed borders would not alter the critical
behavior of avalanches in abelian sandpile models.

\begin{figure}[p]
\begin{center}
\setlength{\unitlength}{1cm}
\begin{picture}(6,8.5)
\put(0.85,6.0){\framebox(1.5,1.5){$S^1\!\!\times\! R$}}
\put(0.85,6.0){\vector(0,1){0.85}}
\put(2.35,6.0){\vector(0,1){0.85}}
\put(0.55,7.6){$\scriptstyle A$}
\put(2.33,7.6){$\scriptstyle B=A$}
\put(0.55,5.7){$\scriptstyle C$}
\put(2.33,5.7){$\scriptstyle D=C$}
\put(1.4,5.6){$\scriptstyle\rm (a)$}
\put(3.65,6.0){\framebox(1.5,1.5){$M$}}
\put(3.65,6.0){\vector(0,1){0.85}}
\put(5.15,7.5){\vector(0,-1){0.85}}
\put(3.35,7.6){$\scriptstyle A$}
\put(5.13,7.6){$\scriptstyle B=C$}
\put(3.35,5.7){$\scriptstyle C$}
\put(5.13,5.7){$\scriptstyle D=A$}
\put(4.2,5.6){$\scriptstyle\rm (b)$}
\put(0.85,3.5){\framebox(1.5,1.5){$S^1\!\!\times\! S^1$}}
\put(0.85,3.5){\vector(0,1){0.85}}
\put(2.35,3.5){\vector(0,1){0.85}}
\put(0.85,3.5){\vector(1,0){0.75}}
\put(0.85,3.5){\vector(1,0){0.85}}
\put(0.85,5.0){\vector(1,0){0.75}}
\put(0.85,5.0){\vector(1,0){0.85}}
\put(0.55,5.1){$\scriptstyle A$}
\put(2.33,5.1){$\scriptstyle B=A$}
\put(0.55,3.2){$\scriptstyle C$}
\put(2.33,3.2){$\scriptstyle D=C$}
\put(1.4,3.1){$\scriptstyle\rm (c)$}
\put(3.65,3.5){\framebox(1.5,1.5){$K$}}
\put(3.65,3.5){\vector(0,1){0.85}}
\put(5.15,5.0){\vector(0,-1){0.85}}
\put(3.65,3.5){\vector(1,0){0.75}}
\put(3.65,3.5){\vector(1,0){0.85}}
\put(3.65,5.0){\vector(1,0){0.75}}
\put(3.655,5.0){\vector(1,0){0.85}}
\put(3.35,5.1){$\scriptstyle A$}
\put(5.13,5.1){$\scriptstyle B=C$}
\put(3.35,3.2){$\scriptstyle C$}
\put(5.13,3.2){$\scriptstyle D=A$}
\put(4.2,3.1){$\scriptstyle\rm (d)$}
\put(0.85,1.0){\framebox(1.5,1.5){$RP$}}
\put(0.85,1.0){\vector(0,1){0.85}}
\put(2.35,2.5){\vector(0,-1){0.85}}
\put(0.85,1.0){\vector(1,0){0.75}}
\put(0.85,1.0){\vector(1,0){0.85}}
\put(2.35,2.5){\vector(-1,0){0.75}}
\put(2.35,2.5){\vector(-1,0){0.85}}
\put(0.55,2.6){$\scriptstyle A$}
\put(2.33,2.6){$\scriptstyle B=C$}
\put(0.55,0.7){$\scriptstyle C$}
\put(2.33,0.7){$\scriptstyle D=A$}
\put(1.4,0.6){$\scriptstyle\rm (e)$}
\put(3.65,1.0){\framebox(1.5,1.5){$S^2$}}
\put(3.65,1.0){\vector(0,1){0.75}}
\put(3.65,1.0){\vector(0,1){0.85}}
\put(3.65,1.0){\vector(1,0){0.75}}
\put(3.65,1.0){\vector(1,0){0.85}}
\put(5.15,1.0){\vector(0,1){0.85}}
\put(3.65,2.5){\vector(1,0){0.85}}

\put(3.35,2.6){$\scriptstyle A$}
\put(5.13,2.6){$\scriptstyle B$}
\put(3.35,0.7){$\scriptstyle C$}
\put(5.13,0.7){$\scriptstyle D=A$}
\put(4.2,0.6){$\scriptstyle\rm (f)$}
\end{picture}
\end{center}
\caption{\label{top}
Possible topologies obtained by imposing boundary conditions
on a two-dimensional lattice. The sides are identified
according to the arrows type and direction. Lattices (a), (c), and (f) are
orientables, while (b), (d), and (e) are not. Only (a) and (b) are
non-compact.}
\end{figure}
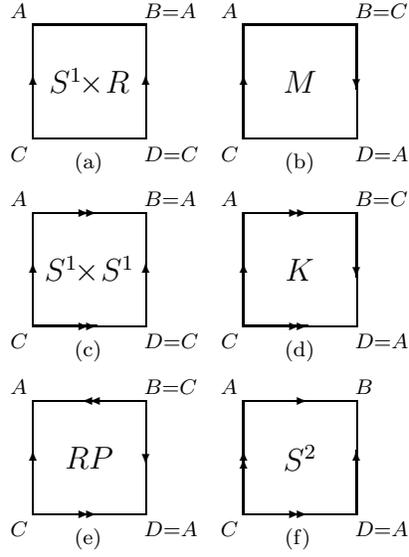

We consider critical height sandpile models. In such models, 
the number of sand grains are given by the
integers $h({\bf r})$ and the 
dynamics is given by the following automaton rule,
valid for any lattice dimension $d$,
\begin{equation}
\label{auto}
{\rm if\ } h({\bf r}') > d-1, {\ \rm then \ }
h({\bf r}) = h({\bf r}) - \Delta_{{\bf r}'}({\bf r}),
\end{equation}
where $\Delta_{{\bf r}'}({\bf r})$ is the Laplacian matrix:
$\Delta_{{\bf r}'}({\bf r}')=2d$, 
$\Delta_{{\bf r}'}({\bf r})=-1$ if $r$ is a nearest neighbor 
of $\bf r'$, and 
$\Delta_{{\bf r}'}({\bf r})=0$ otherwise.
The choice of $k=d-1$ in (\ref{auto}) is not important since a shift in
$k$ simply implies a shift in $h$. The rule  (\ref{auto}) is
repeated until $h\le d-1$ for all points in the lattice. If a sand grain
reaches a boundary $\partial$, 
it shall drop off the edge. Such ``free'' boundary
condition guarantees that the system stops in a minimally stable state.
In order to impose the free boundary conditions, we assume that for
${\bf r}'\in\partial$ the Laplacian matrix is the same one used for the bulk,
although the set of nearest neighbor of ${\bf r}'$ is reduced.
The different lattice topologies are defined from some border identification
of the usual lattice, see Fig. \ref{top}. In our case, they 
are actually imposed in the Laplacian matrix as we will explain bellow. 

We begin with the two-dimensional case. The usual rectangular
lattice with free boundary conditions corresponds to the topology $R^2$.
The cylindrical lattice can be defined by a periodic 
condition on the standard $N\times M$ lattice:
\begin{equation}
\label{cil}
h(a{\bf e}_1 + (b+M){\bf e}_2) = h(a{\bf e}_1 + b{\bf e}_2),
\end{equation}
where ${\bf e}_1$ and ${\bf e}_2$ are local generators of the lattice,
and $1\le a \le N$ and $1\le b \le M$ are integers. On the other
hand, the M\"obius-strip topology is defined by
\begin{equation}
\label{mob}
h(a{\bf e}_1 + (b+M){\bf e}_2) = h((N+1-a){\bf e}_1 + b{\bf e}_2).
\end{equation}
In the first case, if a sand grain reaches the point $(a, M+1)$
it will appear in $(a,1)$, and in the second one it would appear
in $(N+1-a,1)$. In the same way,
if it reaches the point $(a,0)$ it will appear in $(a,M)$ for
the cylindrical lattice and in $(N+1-a,M)$ for the M\"obius-strip.
Such conditions can be easily implemented by means of the Laplacian
matrix. For the cylindrical case, for instance, we need only to define
as nearest neighbor of the point $a{\bf e}_1 + M{\bf e}_2$,
$1<a<N$, the
four points 
$(a\pm 1){\bf e}_1 + (M-1){\bf e}_2$ and 
$(a\pm 1){\bf e}_1 + {\bf e}_2$. Analogously, the nearest neighbor
of $a{\bf e}_1 + {\bf e}_2$ would be
$(a\pm 1){\bf e}_1 + 2{\bf e}_2$ and 
$(a\pm 1){\bf e}_1 + M{\bf e}_2$. 
In two dimensions one does not have others topologically
inequivalent lattices with relevance to our purposes. The convergence
of (\ref{auto}) requires free boundary conditions
and thus the lattice cannot be compact. The only non-compact lattices
that can be constructed by imposing boundary conditions on
the standard two-dimensional 
lattice are the cylindrical and the M\"obius-strip ones, see Fig. \ref{top}.
For the three-dimensional case we are also restricted to 
non-compact lattices to guarantee the convergence of (\ref{auto}).
The non-compact three-dimensional lattice that
we examine are: $R^3$,
$S^1\times R^2$, $M\times R$, $S^1\times S^1\times R$, 
 $S^2\times R$,
$K\times R$, and $RP\times R$. The first three topologies are straightforward
extensions of the two-dimensional cases and the last ones are obtained by
imposing boundary conditions as follows. For an 
$L\times N \times M$ lattice $a{\bf e}_1 + b{\bf e}_2 + c{\bf e}_3$,
we define $L$ two-dimensional sections by fixing $a$ and considering
$b$ and $c$ as free. For each one of these sections we 
define the Laplacian matrix according to Fig. \ref{top},
imposing the boundary conditions  
in an analogous way of (\ref{cil}) and (\ref{mob}). Obviously, the
$S^2\times R$ topology requires $N=M$. 

\begin{figure}[p]
\hfill\hbox{\epsfxsize=10cm\epsfbox{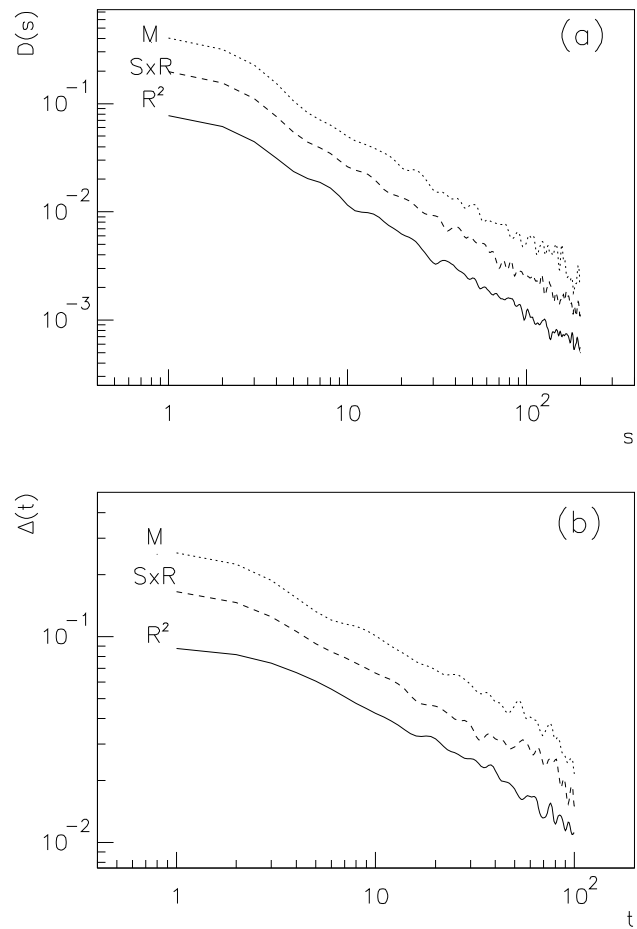}}\hfill\hfill
\caption{\label{two}
Two-dimensional lattices: (a) Distribution of cluster sizes, 
(b) Weighted distribution of fluctuation lifetimes.
Results for $100\times 100$ lattices, each topology is averaged over
400 samples.}
\end{figure}

\begin{figure}[p]
\hfill\hbox{\epsfxsize=10cm\epsfbox{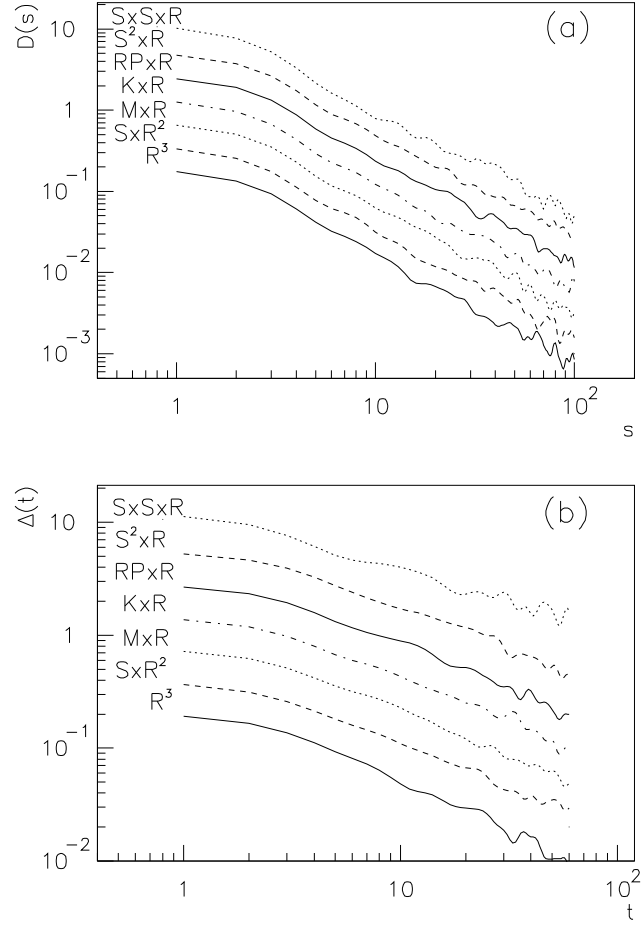}}\hfill\hfill
\caption{\label{three}
Three-dimensional lattices: (a) Distribution of cluster sizes, 
(b) Weighted distribution of fluctuation lifetimes.
Results for $30\times 30\times 30$ lattices, each topology is averaged over
400 samples.}
\end{figure}

In our simulations we 
start with a void lattice $(h({\bf r})=0)$,  choose
at random a point ${\bf r}'$ and set 
$h({\bf r}')\gg 1$. The system then evolves following 
(\ref{auto}) until it reaches a stable state. This state is then locally
perturbated by adding a sand grain in an aleatory point 
${\bf r}''$. If the stability is broken, the system will
evolve and we measure the ``cluster size'' $s$, given by the number 
of sites toppling at least once, the total number of topplings $m$, 
and the ``fluctuation lifetime'', given by the number
of iteration $t$ necessary to reach again a stable state.
The distribution of cluster sizes, of toppling numbers,
and the distribution
of lifetimes weighted by the temporal average of the response $(s/t)$
obey respectively the following power-laws 
\begin{eqnarray}
D(s) &\propto& s^{-\tau}, \nonumber \\
D(m) &\propto& m^{-\tau_m}, \nonumber \\
\Delta(t) = \frac{s}{t}D(t) &\propto& t^{-b}.
\end{eqnarray}
The Table \ref{t1} presents the values of the critical exponents
for the various topologies.
Figures \ref{two} and 
\ref{three} show the distribution of cluster sizes and the
weighted distribution of
lifetimes for  $100\times 100$ and $30\times 30\times 30$ lattices
respectively. 
The distributions are averaged
for each topology over 400 distinct samples. 
The situation is unchanged for lattices of different sizes.
Our values for the spatial critical exponents ($\tau$ and $\tau_m$)
are in good agreement 
with the original results for two- and three-dimensional 
critical height models\cite{BTW,BT}.   
For all distributions, the curves
are linear over some decades but deviate for small and for large
values of $s$, $m$, and $t$. As it is well known in the literature, the first
can be understood as effect of the discreteness of the lattice and the
second as finite-size effects. Also, the distribution
of lifetimes gives less impressive results because typically we have 
$t\ll s\approx m$, reducing considerably the number of events.
However, we do not attribute the discrepancies in the $b$-exponent
for three-dimensional lattices to such a
fact, we instead attribute them to genuine dependence
on the lattice topology.

Both set results are compatible with the condition $\tau=\tau_m$. Such
condition was first noticed in the numerical analysis of \cite{M} and
it was also considered recently in \cite{STC}. We notice also that 
our results agree with the 
theoretical predictions for $\tau$ and $b$
of Zhang's continuous-energy model\cite{Z},
according to which $\tau_{\rm T}=1$ and $b_{\rm T}=1/2$ for $d=2$, and
$\tau_{\rm T}=4/3$  for $d=3$, 
although Zhang's model are based in usual hyper-cubic lattices.

We finish noticing that, as it was already said, the distributions
characterizing border avalanches do depend on the lattice topology.
We denote by $f$ the number of sand grains that drop off the edges. By
applying standard finite-size scaling analysis, we fit the $f$-distribution
$D(f,L)$ for a lattice of linear size $L$ as
\begin{equation}
D(f,L) =  L^{-\beta} g(f/L^\nu), {\rm\ for\ } f,L\gg 1,
\end{equation}
where $\beta$ and $\nu$ are the critical indices and $g$ is the scalling
function. We could verify that both critical indices and the scalling
function do depend on the lattice topology.

To summarize, we have demonstrated by numerical simulations 
for two- and three-dimensional critical height sandpile
models  that the spatial distributions characterizing
bulk avalanches do not depend on the lattice topology. 
For the temporal distribution, only for the three-dimensional case
there is a dependence of the topology. The 
theoretical prediction of Zhang's continuous-energy model\cite{Z},
$\tau_{\rm T} = 2-2/d$, 
seems to be valid also with non-trivial lattice 
topologies, in spite of Zhang's original calculations were made 
exclusively for $R^d$. It would be worth to understand 
analytically such a coincidence. Simulations on huge lattices, in
the line of \cite{M} for instance, would be  of great interest to
improve the data for the evaluation of the $b$-exponent for three-dimensional
lattices.  

This work was supported by CNPq, under grant 201630/93-1. 
N. Armesto and G. Parente are acknowledged for helpful discussions.

\begin{table}[p]
\begin{center}
\begin{minipage}{6.5cm}
\begin{tabular}{| l | c | c | c |} 
                     & $\tau$ & $\tau_m$ & $b$   \\ \hline
 $R^2$               &$1.02$ &$1.02$&$0.49$\\ \hline
 $S^1\times R$       &$1.04$ &$1.02$&$0.52$\\ \hline
 $M\times R$         &$0.99$ &$1.00$&$0.50$\\ \hline\hline

 $R^3$                   &$1.31$ &$1.31$&$0.88$\\ \hline
 $S^1\times R^2$         &$1.30$ &$1.31$&$0.78$\\ \hline
 $S^1\times S^1\times R$ &$1.31$ &$1.31$&$0.62$\\ \hline
 $M\times R$             &$1.29$ &$1.30$&$0.79$\\ \hline 
 $S^2\times R$           &$1.30$ &$1.29$&$0.73$\\ \hline
 $RP\times R$            &$1.29$ &$1.31$&$0.77$\\ \hline
 $K\times R$             &$1.32$ &$1.32$&$0.75$\\ 
\end{tabular}
\end{minipage}
\end{center}
\caption{\label{t1} 
Critical exponents for avalanches in critical height sandpile models based in
 two- and three-dimensional lattices of various topologies. All exponents
have an accuracy of $\pm 0.01$.}
\end{table}

\begin{references}


\bibitem{BTW} P. Bak, C. Tang, and K. Wiesenfeld, Phys. Rev. Lett. {\bf 59},
381 (1987); Phys. Rev. {\bf A38}, 364 (1988).
\bibitem{N}S.R. Nagel, Rev. Mod. Phys. {\bf 64}, 321 (1992).
\bibitem{P}H. Puhl, Physica {\bf A182}, 295 (1992).
\bibitem{BT}  C. Tang and  P. Bak, Phys. Rev. Lett. {\bf 60}, 2347 (1988).
\bibitem{Z}Y.-C. Zhang, Phys. Rev. Lett. {\bf 63}, 470 (1989).
\bibitem{an}D. Dhar and  R. Ramaswamy, Phys. Rev. Lett. {\bf 63}, 1659 (1989);
D. Dhar, Phys. Rev. Lett. {\bf 64}, 1613 (1990); P. Grassberger and S.S. Manna,
J. Phys. (France) {\bf 51}, 1077 (1990); S.S. Manna,
 Physica {\bf A179}, 249 (1991); 
S.N. Majumdar and D. Dhar,  Physica {\bf A185}, 129 (1992);
E.V. Ivashkevich, J. Phys. {\bf A27}, 3643 (1994); 
S.W. Chan and H.F. Chau, Physica {\bf A216}, 227 (1995).
\bibitem{DH}P. Datta and P.M. Horn, Rev. Mod. Phys. {\bf 53}, 497 (1981).
\bibitem{KNWZ} L.P. Kadanoff, S.R. Nagel, L. Wu, and S. Zhou,
Phys. Rev. {\bf A39}, 6524 (1989).
\bibitem{O} C.-H. Lu, H.M. Jaeger, and S.R. Nagel,
Phys. Rev. {\bf A43}, 7091 (1991); B. Tadic, U. Nowak, K.D. Usadel,
R. Ramaswamy, S. Padlewski, Phys. Rev. {\bf A45}, 8536 (1992).
\bibitem{STC}A.L. Stella, C. Tebaldi, and G. Caldarelli, 
Phys. Rev. {\bf E52}, 72 (1995).
\bibitem{P1}H. Puhl,  Physica {\bf A197}, 14 (1993).
\bibitem{D}J.A.M.S. Duarte and N.J.A.P. Gon\c calves,
 Physica {\bf A168}, 901 (1990).
\bibitem{naka} M. Nakahara, {\em Geometry, Topology and Physics},
Adam Hilger, Bristol, 1990.
\bibitem{M} S.S. Manna, J. Stat. Phys. {\bf 59}, 509 (1990).
\end{references}
\end{document}